1
  2
  3
  4
  5
  6
  7
  8
  9
 10
 11
 12
 13
 14
 15
 16
 17
 18
 19
 20
 21
 22
 23
 24
 25
 26
 27
 28
 29
 30
 31
 32
 33
 34
 35
 36
 37
 38
 39
 40
 41
 42
 43
 44
 45
 46
 47
 48
 49
 50
 51
 52
 53
 54
 55
 56
 57
 58
 59
 60
 61
 62
 63
 64
 65
 66
 67
 68
 69
 70
 71
 72
 73
 74
 75
 76
 77
 78
 79
 80
 81
 82
 83
 84
 85
 86
 87
 88
 89
 90
 91
 92
 93
 94
 95
 96
 97
 98
 99
100
101
102
103
104
105
106
107
108
109
110
111
112
113
114
115
116
117
118
119
120
121
122
123
124
125
126
127
128
129
130
131
132
133
134
135
136
137
138
139
140
141
142
143
144
145
146
147
148
149
150
151
152
153
154
155
156
157
158
159
160
161
162
163
164
165
166
167
168
169
170
171
172
173
174
175
176
177
178
179
180
181
182
183
184
185
186
187
188
189
190
191
192
193
194
195
196
197
198
199
200
201
202
203
204
205
206
207
208
209
210
211
212
213
214
215
216
217
218
219
220
221
222
223
224
225
226
227
228
229
230
231
232
233
234
235
236
237
238
239
240
241
242
243
244
245
246
247
248
249
250
251
252
253
254
255
256
257
258
259
260
261
262
263
264
265
266
267
268
269
270
271
272
273
274
275
276
277
278
279
280
281
282
283
284
285
286
287
288
289
290
291
292
293
294
295
296
297
298
299
300
301
302
303
304
305
306
307
308
309
310
311
312
313
314
315
316
317
318
319
320
321
322
323
324
325
326
327
328
329
330
331
332
333
334
335
336
337
338
339
340
341
342
343
344
345
346
347
348
349
350
351
352
353
354
355
356
357
358
359
360
361
362
363
364
365
366
367
368
369
370
371
372
373
374
375
376
377
378
379
380
381
382
383
384
385
386
387
388
389
390
391
392
393
394
395
396
397
398
399
400
401
402
403
404
405
406
407
408
409
410
411
412
413
414
415
416
417
418
419
420
421
422
423
424
425
426
427
428
429
430
431
432
433
434
435
436
437
438
439
440
441
442
443
444
445
446
447
448
449
450
451
452
453
454
455
456
457
458
459
460
461
462
463
464
465
466
467
468
469
470
471
472
473
474
475
476
477
478
479
480
481
482
483
484
485
486
487
488
489
490
491
492
493
494
495
496
497
498
499
500
501
502
503
504
505
506
507
508
509
510
511
512
513
514
515
516
517
518
519
520
521
522
523
524
525
526
527
528
529
530
531
532
533
534
535
536
537
538
539
540
541
542
543
544
545
546
547
548
549
550
551
552
553
554
555
556
557
558
559
560
561
562
563
564
565
566
567
568
569
570
571
572
573
574
575
576
577
578
579
580
581
582
583
584
585
586
587
588
589
590
591
592
593
594
595
596
597
598
599
600
601
602
603
604
605
606
607
608
609
610
611
612
613
614
615
616
617
618
619
620
621
622
623
624
625
626
627
628
629
630
631
632
633
634
635
636
637
638
639
640
641
642
643
644
645
646
647
648
649
650
651
652
653
654
655
656
657
658
659
660
661
662
663
664
665
666
667
668
669
670
671
672
673
674
675
676
677
678
679
680
681
682
683
684
685
686
687
688
689
690
691
692
693
694
695
696
697
698
699
700
701
702
703
704
705
706
707
708
709
710
711
712
713
714
715
716
717
718
719
720
721
722
723
724
725
726
727
728
729
730
731
732
733
734
735
736
737
738
739
740
741
742
743
744
745
746
747
748
749
750
751
752
753
754
755
756
757
758
759
760
761
762
763
764
765
766
767
768
769
770
771
772
773
774
775
776
777
778
779
780
781
782
783
784
785
786
787
788
789
790
791
792
793
794
795
796
797
798
799
800
801
802
803
804
805
806
807
808
809
810
811
\documentclass[prd,preprint,superscriptaddress,amsmath,amssymb,nofootinbib]{revtex4}
\usepackage{graphicx,color,slashed}% Include figure files

\def\be{\begin{eqnarray}}
\def\ed{\end{eqnarray}}

\begin{document}

%{\begin{flushright}{xxx}
%\end{flushright}}

\title{\bf \Large The Process $gg\to h^0 Z^{*}$ in the Inverted Hierarchy Scenario of the 2HDM Type-I at the LHC}

\author{A.G. Akeroyd}
\email{a.g.akeroyd@soton.ac.uk}
\affiliation{School of Physics and Astronomy, University of Southampton,
Highfield, Southampton SO17 1BJ, United Kingdom}

\author{S. Alanazi}
\email{swa1a19@soton.ac.uk; SWAlanazi@imamu.edu.sa}
\affiliation{School of Physics and Astronomy, University of Southampton,
Highfield, Southampton SO17 1BJ, United Kingdom}
\affiliation{Physics Department, Imam Mohammad Ibn Saud Islamic University (IMISU), P.O. Box 90950, Riyadh, 11623, Saudi Arabia}

\author{S. Moretti}
\email{s.moretti@soton.ac.uk; stefano.moretti@physics.uu.se}
\affiliation{School of Physics and Astronomy, University of Southampton,
Highfield, Southampton SO17 1BJ, United Kingdom}
\affiliation{Department of Physics and Astronomy, Uppsala University, Box 516, SE-751 20 Uppsala, Sweden}


\begin{abstract}
{\noindent\footnotesize
While searching at the Large Hadron Collider (LHC) for the production and decay of the CP-odd scalar ($A^0$) in the 2-Higgs-Doublet Model (2HDM) with Natural Flavour Conservation (NFC) via the channels $gg\to A^0$ (through one-loop triangle diagrams) and 
$A^0\to h^0 Z^*$ (with $m_{h^0} =125$ GeV or $m_{h^0} < 125$ GeV, with $Z$ off-shell), respectively, a factorisation of the two processes is normally performed, with the $A^0$ state  being on-shell. While this approach is gauge-invariant, it is not capturing the presence of
either of the following two channels:  $gg\to Z^*\to h^0Z^*$ (through one-loop triangle diagrams) or $gg\to h^0Z^*$ (through one-loop box diagrams). As  the resolution of the $A^0$ mass cannot be infinitely precise, we affirm that all such contributions should be computed simultaneously, whichever the $h^0$($Z^{*}$) decay(splitting) products,
thereby including all possible interferences amongst themselves. The cross section of the ensuing complete process is significantly different from that obtained in the factorisation case, being of the order up to ten percent in either direction at the integrated level and larger (including changes in the shape of kinematical observables) at the differential level. We thus suggest that the complete calculation ought to be performed while searching for $A^0$ in this channel.   We  illustrate this need for the case of a 2HDM of Type-I in the inverted hierarchy scenario with $m_{h^0}<125$ GeV.}
\end{abstract}

\maketitle

\section{Introduction}
\label{Sec:Intro}
\noindent 
While the discovery in 2012 of a neutral spin-0 object, consistent with the Higgs boson predicted within the Standard Model (SM), by the   ATLAS and CMS collaborations of the LHC
\cite{Aad:2012tfa,Chatrchyan:2012xdj} has signified the consolidation of this theoretical framework, despite the many measurements of its properties (mass, width, spin, charge, CP quantum numbers, etc.), it has also done very little to clarify what Beyond the SM (BSM) physics may exist in Nature. In fact, we know that  some form of BSM physics must exist to remedy both theoretical (e.g., the hierarchy problem) and experimental (e.g., neutrino masses, dark matter, baryonic asymmetry of the universe) flaws of the SM.

Indeed, whether or not the observed new boson (with a measured mass of about 125 GeV) is the 
(lone) Higgs state of the SM or else the first manifestation of an enlarged Higgs sector is still an issue that needs to be clarified. In 
particular, it is possible that such a particle belongs to a 2HDM
 \cite{Lee:1973iz,Gunion:1989we,Branco:2011iw,Wang:2022yhm}, in which the scalar potential 
contains two $SU(2)_L\otimes U(1)_Y$ isospin doublets instead of just one like in the SM. Herein,  there exists a so-called ``alignment limit'', wherein  one of the CP-even scalars (of the two predicted) 
 has properties that exactly match those of the Higgs boson of the SM. This condition is naturally obtained
if all other (neutral) Higgs states have masses that are much larger than 125 GeV (the so-called ``alignment with decoupling'' limit). Such an alignment condition, though, can also be realised if
all such (neutral) Higgs states have masses of the order of the Electro-Weak (EW) scale (the so-called ``alignment without decoupling" limit).
The latter will be the focus of this work.

In a 2HDM there are two CP-even (scalars) states, $h^0$ and $H^0$ (with $m_{h^0} < m_{H^0}$), a charge conjugate pair of charged states, $H^+$ and $H^-$, and
a neutral CP-odd (pseudoscalar) state, $A^0$. The
 discovered 125 GeV boson has been shown to be neutral and CP-even. Consequently, in the context of the 2HDM, it could be interpreted as being either $h^0$ (a Higgs mass configuration called ``normal hierarchy" (NH)) or $H^0$ (a Higgs mass configuration called ``inverted hierarchy" (IH)). This work will consider the latter scenario.

The CP-odd $A^0$ does not have tree-level couplings to the
 gauge bosons of the weak interactions $(W^\pm, Z$) and is thus 
inconsistent with the measured properties of the discovered Higgs state. However, it could be the mediator of a signal which may reveal the IH scenario, as it can interact with both the $h^0$ and $H^0$ states, via the production and decay processes $gg\to A^0\to h^0Z^*$ and 
$gg\to A^0\to H^0Z^*$, respectively, where the neutral massive EW gauge boson can be considerably off-shell.
% thereby enabling essentially any mass relation between the $A^0$ state the $h^0$ or $H^0$. 
In the IH scenario, the cross section for the first process is naturally unsuppressed as it is proportional to $|\cos(\beta-\alpha)|^2$, where $\alpha$ is the mixing angle in the CP-even neutral Higgs sector while $\beta$ is the arc tangent of the ratio of the Vacuum Expectation Values (VEVs) of the two Higgs doublets, which is the same coupling strength entering the $H^0VV$ (where $V=W^\pm, Z$) vertex, which has been measured to be close to 1.  
The second process is naturally small, as it is proportional to $|\sin(\beta-\alpha)|^2$. Therefore, it is entirely possible that $gg\to A^0\to h^0Z^*$ is discovered before $gg\to A^0\to H^0Z^*$, leading to the simultaneous discovery of two new Higgs states 
($A^0$ and $h^0$).

We shall therefore focus on the prospects of discovering an $A^0$ from the 2HDM at the LHC via its production and decay process $gg\to A^0\to h^0Z^{*}$ in the context of IH but, unlike most literature, we will refrain from computing such a process on its own with the $A^0$ state treated in Narrow Width Approximation (NWA). Instead, we compute the full $gg\to h^0Z^*$ process, wherein, alongside the above channel (but modelled using a finite value of $\Gamma_{A^0}$), we also consider  $gg\to Z^*\to h^0Z^{*}$ and all other diagrams entering (gauge-invariantly) the process $gg\to h^0Z^{*}$, including all interferences. We shall see that such a treatment will lead to a non-negligible modification of the sensitivity obtained using the $A^0$ diagram (only) in NWA.  Thus, our present work expands and surpasses what we did in Ref.~\cite{Akeroyd:2023kek}, where we studied the factorised process in NWA. Also, our results are corroborated by what some of us noted in Ref.~\cite{Accomando:2020vbo}, where effects of the additional topologies, with respect to that of the factorised process, were found to be sizeable, although that paper was concerned with a NH scenario in a so-called 2HDM Type-II, whereas we will be studying here  a 2HDM Type-I in IH. (See later on for a definition of (Yukawa) `Types' in the context of the 2HDM.) 

This work is organised as follows. In Sec.~\ref{Sec:2HDM}  we introduce the  2HDM that we will be using. In Sec.~\ref{Sec:Parameters}
we discuss its available parameter space in the light of current theoretical and experimetal constraints. 
 Then, in Sec.~\ref{Sec:Results}, we present our results. Finally, our conclusions are given in Sec.~\ref{Sec:Summary}.

\section{The 2HDM}
\label{Sec:2HDM}
\noindent
The SM embeds one complex scalar isospin doublet $(I=1/2)$ with hypercharge $Y=1$, in which the real part of the neutral scalar field obtains a VEV, denoted by $v$, which is approximately 246 GeV.
The presence of such a non-zero value of $v$ in the Lagrangian leads to the spontaneous breaking of the $SU(2)_L\otimes U(1)_Y$ local gauge symmetry to a $U(1)_Q$ one, in turn providing mass to the $W^\pm$ and $ Z$ gauge bosons (via the kinetic energy terms of the (pseudo)scalar fields)
and charged fermions (via the Yukawa couplings). Such an EW Symmetry Breaking (EWSB) pattern is triggered by the so-called  ``Higgs mechanism". 
In the context of the SM, wherein the Higgs mechanism is implemented through a single Higgs doublet field, a single physical Higgs state is predicted, $h_{\rm SM}$, which can be identified with 
the particle discovered at the LHC in 2012, with mass of about 125 GeV. Thus, of the 4 degrees of freedom associated with the SM Higgs doublet field, 3 are used to generate a longitudinal polarisation for the aforementioned $W^\pm$ and $Z$ states (indeed, only possible for massive spin-1 gauge fields).
However,  the Higgs mechanism can also be implemented using 2 complex scalar doublets for which there are now two VEVs ($v_1$ and $v_2$) and such a scenario is called the 2HDM \cite{Lee:1973iz,Gunion:1989we,Branco:2011iw,Wang:2022yhm}.
Herein, after EWSB (which again leads to massive $W^\pm$ and $Z$ gauge bosons,) 
there are 5 physical Higgs bosons remaining, as mentioned in the Introduction.

 In the context of the 2HDM, the 125 GeV boson discovered at the LHC is interpreted as being either
  $h^0$ (NH) or $H^0$ (IH), with couplings very close to those of the SM Higgs boson. The second configuration leads to interesting new phenomenology, as there would be at least one Higgs state lighter than 125 GeV. 
However, enlarging the scalar sector of the SM can conflict with experimental data. Firstly, a strong suppression from Flavour Changing Neutral Currents (FCNCs) data implies 
 stringent constraints
on the Yukawa structure of the 2HDM.
In fact, the Yukawa couplings in the 2HDM are not flavour diagonal, in turn leading to potentially large Higgs mediated FCNCs, 
which must then be  suppressed.
A particularly elegant suppression mechanism of FCNCs in the 2HDM is invoking NFC, by exploiting the  
Paschos-Glashow-Weinberg theorem \cite{Glashow:1976nt}, requiring that the Lagrangian respects certain discrete  ($Z_2$) symmetries.
Such symmetries enforce that a given flavour of charged fermion receives its mass from just one VEV, in turn leading
to the elimination of FCNC processes at the tree level. In reality, such a $Z_2$ symmetry can be softly broken  while still complying with the aforementioned data:  this allows for more freedom in achieving EWSB.

The  most general scalar potential of a 2HDM that is invariant under the $SU(2)_L\otimes U(1)_Y$ local gauge symmetry and which breaks (via the $m^2_{12}$ term) the $Z_2$ symmetry only softly is written as  \cite{Gunion:1989we,Branco:2011iw}:
 \begin{eqnarray}
 %\begin{align}
        V(\Phi _{1}\Phi _{2})  =  m_{11}^{2}\Phi _{1}^{\dagger }\Phi _{1}+m_{22}^{2}\Phi _{2}^{\dagger }\Phi _{2}-  m_{12}^{2}(\Phi _{1}^{\dagger }\Phi _{2}+\Phi _{2}^{\dagger }\Phi _{1})+  \frac{ \lambda _{1}}{2}(\Phi _{1}^{\dagger }\Phi _{1})^{2}+\\ \nonumber
         \frac{ \lambda _{2}}{2}(\Phi _{2}^{\dagger }\Phi _{2})^{2}+  \lambda _{3}\Phi _{1}^{\dagger }\Phi _{1}\Phi _{2}^{\dagger }\Phi _{2}+
          \lambda _{4}\Phi _{1}^{\dagger }\Phi _{2}\Phi _{2}^{\dagger }\Phi _{1}+\frac{ \lambda _{5}}{2}[(\Phi _{1}^{\dagger }\Phi _{2})^{2}+(\Phi _{2}^{\dagger }\Phi _{1})^{2}]\,,
 %\end{align}
\label{Pot}
 \end{eqnarray}
with $\Phi _{i}=\binom{\Phi _{i}^{\dotplus }}{\frac{(\upsilon _{i}+\rho _{i}+i\eta _{i})}{\sqrt{2}}}  \:{\rm and} \; i=1,2$. \\
In general, some of the parameters in such a potential can be complex and thus they can generate CP violating effects, both spontaneously (via complex VEVs) or explicitly (through the coefficients of the operators in eq.~(\ref{Pot})). Here, we consider a simplified scenario by taking all such parameters to be real, as is often done
in phenomenological studies of the 2HDM.
The scalar potential then has 8 real independent parameters: $m^2_{11}$, $m^2_{22}$, $m^2_{12}$, $\lambda _{1}$, $\lambda _{2}$, $\lambda _{3}$, $\lambda _{4}$ and $\lambda _{5}$.
These parameters determine the masses of the Higgs bosons and their couplings to fermions, gauge bosons and amongst  themselves. However, it is convenient to work with different independent parameters that are more directly related to physical observables.
A common choice is: $m_{h^0}$, $m_{H^0}$, $m_{H^\pm}$, $m_{A^0}$, $\upsilon_1$, $\upsilon_2$, $m^2_{12}$ and $\sin(\beta-\alpha)$. The first four parameters are the masses of the physical Higgs bosons.
The VEVs $\upsilon_1$ and $\upsilon_2$ are the values of the neutral CP-even fields in $\Phi_1$ and $\Phi_2$, respectively, at the minimum of the scalar potential:
\begin{equation}
    \left<\Phi _{1} \right> =\frac{1}{\sqrt{2}}\binom{0}{\upsilon _{1}}\, ,\, \, \, \, \left<\Phi _{2} \right> =\frac{1}{\sqrt{2}}\binom{0}{\upsilon _{2}}\,.\\[0.15cm]
\end{equation}
The parameter $\beta$ is defined via $\tan\beta=\upsilon_2/\upsilon_1$ while the angle $\alpha$ determines the composition of the CP-even mass eigenstates $h^0$ and $H^0$ in terms of the original 
neutral CP-even fields that are present in the isospin doublets $\Phi_1$ and $\Phi_2$. Of these 8 parameters in the scalar potential, 2 have now been measured. Firstly, after EWSB in the 2HDM, the mass of the $W^\pm$ boson is given by
$m_W=gv/2$, with $ \upsilon=\sqrt{\upsilon^{2}_{1}+\upsilon^{2}_{2}}\simeq 246$ GeV, hence, only one of $\upsilon_1$ and $\upsilon_2$ is independent ($\tan\beta=\upsilon_2/\upsilon_1$ is therefore taken as an independent parameter). Secondly, in the 2HDM, the discovered 125 GeV boson is taken to be $h^0$ or $H^0$ and thus either $m_{h^0}=125$ GeV (NH) or $m_{H^0}=125$ GeV (IH) is fixed.
The remaining 6 independent parameters in the 2HDM scalar potential are therefore: $m_{H^\pm}$, $m_{A^0}$, $m^2_{12}$, $\tan\beta$, $\sin(\beta-\alpha)$ and one of $[m_{h^0}, m_{H^0}]$. In the NH scenario $m_{H^0}>125$ GeV
and in the IH scenario $m_{h^0}<125$ GeV. As intimated, in this work we shall be focussing on the IH scenario and study the phenomenology of the $A^0$ and $h^0$ states in relation to each other.

As mentioned above, the  masses of the pseudoscalar $A^0$ and charged $H^{\pm}$ are independent inputs parameters. In terms of the original parameters in the
scalar potential, these masses are given by:
   \begin{equation}
       \begin{aligned}
         m_{A^0}^{2} & = \left [\frac{ m_{12}^{2}}{\upsilon _{1}\upsilon _{2}}  -2\lambda_{5}\right ](\upsilon _{1}^{2}+\upsilon _{2}^{2})\,,\\ m_{H^{\pm }}^{2} & = \left [\frac{ m_{12}^{2}}{\upsilon _{1}\upsilon _{2}} -\lambda_{4} -\lambda_{5}\right ](\upsilon _{1}^{2}+\upsilon _{2}^{2}) = \left [ m_{A}^{2}  +\upsilon (\lambda_{5}-\lambda_{4})\right ]  \,.
       \end{aligned}
   \end{equation}
  From these equations, it can be seen that the mass difference $m_{A^0}-m_{H^\pm}$ depends on $\lambda_5-\lambda_4$. In our numerical analysis,  we
  shall be taking $m_{A^0}=m_{H^\pm}$ in order to satisfy  easily the constraints from EW Precision Observables (EWPOs),  including the so-called `oblique parameters', which corresponds to $\lambda_5=\lambda_4$, and scan these two (equal) masses between 140 and 170 GeV.
  For the masses of the CP-even scalars, as mentioned earlier, we take $m_{H^0}=125$ GeV and  {{ $ m_{h^0} < 125$ GeV}} (IH scenario).  
     
  There are four distinct types of 2HDM with NFC which differ in how the two doublets are coupled to the charged fermions. These choices are referred to as follows: Type-I, Type-II, Lepton Specific and Flipped \cite{Barger}.
% The phenomenology of all four models has been studied in great detail.
 The Lagrangian terms  in the 2HDM that describe the Yukawa  interactions of $A^0$ with 
the fermions can be written as  \cite{Branco:2011iw}:
\begin{equation}
{\cal L}^{\rm Yukawa}_{A^0} =\frac{i}{v}\left(y^d_{A^0} m_d A^0 \overline d \gamma_5 d + y^u_{A^0} m_u A^0 \overline u \gamma_5 u+  y^l_{A^0} m_l  A^0 \overline{l} \gamma_5 l \right)\,.
\label{yukawa}
\end{equation}
In eq.~(\ref{yukawa}),  $d(u)[l]$ refers to the down(up)-type quarks[leptons],  i.e., there are three terms of the form $y^d_{A^0} m_d \overline d \gamma_5 d$.
 In Tab.~\ref{2HDMAcoup}, the couplings $y^d_{A^0}$, $y^u_{A^0}$ and $y^l_{A^0}$  of the $A^0$ state to the charged fermions in each of the four Types
are displayed.
\begin{table}[h]
\begin{center}
\begin{tabular}{|c||c|c|c|}
\hline
& $y^d_{A^0}$ &  $y^u_{A^0}$ &  $y^l_{A^0}$ \\ \hline
Type-I
&  $-\cot\beta$ & $\cot\beta$ & $-\cot\beta$ \\
Type-II
& $\tan\beta$ & $\cot\beta$ & $\tan\beta$ \\
Lepton Specific
& $-\cot\beta$ & $\cot\beta$ & $\tan\beta$ \\
Flipped
& $\tan\beta$ & $\cot\beta$ & $-\cot\beta$ \\
\hline
\end{tabular}
\end{center}
\caption{The couplings $y^d_{A^0}$, $y^u_{A^0}$, and $y^l_{A^0}$  in the Yukawa interactions of $A^0$ in the four versions of the 2HDM with NFC.}
\label{2HDMAcoup}
\end{table}

\section{Parameter Space}
\label{Sec:Parameters}
The viable parameter space in a 2HDM must respect all theoretical and experimental constraints, which are listed below.
%The theoretical 
%constraints are: \\ i) vacuum stability, ii) unitarity, iii) perturbativity. The experimental constraints are: 
%) constraints from searching for Higgs bosons of the 2HDM  at past and present colliders  (e.g., LEP, Tevatron and LHC),
%ii) constraints from electroweak precision measurements.
\begin{enumerate}
    \item \underbar{Theoretical constraints}
    \begin{enumerate}
        \item[{(i)}] Vacuum stability\\
         The values of $\lambda_i$ are constrained by the requirement
        that the scalar potential: a) breaks the EW symmetry $SU(2)_L\otimes U(1)_Y$
        to $U(1)_Q$, b) the scalar potential is bounded from below  and c) the scalar potential  stays positive for arbitrarily large values of the scalar fields.  The constraints are: 
         $\lambda _{1}> 0$, $\lambda _{2}> 0$,  $\lambda _{3}+\lambda _{4}-\left|\lambda _{5} \right| + \sqrt{\lambda _{1} \lambda _{2}} \ge 0, \;\;\lambda _{3}+ \sqrt{\lambda _{1}\lambda _{2}}\ge 0$.\\
         From these conditions it can be seen that $\lambda_1$ and $\lambda_2$ are positive definite while $\lambda_3,  \lambda_4$ and $\lambda_5$ can have either sign. 
        \item[{(ii)}] Perturbativity\\
         For calculational purposes it is required that the quartic couplings $\lambda _{i}$ do not take numerical values for which the perturbative expansion  ceases to converge.
        The couplings $\lambda _{i}$ remain perturbative up to the unification scale if 
        they satisfy the condition $\left| \lambda  _{i}\right|\leq 8\pi $.   
        \item[{(iii)}]  Unitarity \\
        The  $2\to 2$ scattering processes ($s_1s_2\to s_3s_4$) involving only (pseudo)scalars $s_i$ (including Goldstone bosons in the generic $R_\xi$ gauge) are mediated by scalar quartic couplings, which depend on the parameters of the scalar potential. 
        Tree-level unitarity constraints require that the eigenvalues of the scattering matrix of the amplitudes for $s_1s_2\to s_3s_4$ be less than the unitarity limit of $8\pi$, which leads to further constraints on $\lambda_i$.
    \end{enumerate}
    \item \underbar{Experimental constraints} 
     \begin{enumerate}
     \item[{(i)}] 
    Direct searches for Higgs bosons\\
    The observation of the 125 GeV boson at the LHC and the non-observation of additional Higgs bosons at LEP, Tevatron and LHC rule out regions of the parameter space of a 2HDM.    
    In our numerical results, these constraints are respected by using the publicly available codes HiggsBounds \cite{Bechtle:2020pkv} (which implements searches for additional Higgs bosons) and HiggsSignals \cite{Bechtle:2020uwn}
    (which implements the measurements of the 125 GeV boson). Any point in the 2HDM 
    parameter space that violates experimental limits/measurements concerning Higgs bosons is rejected\footnote{We have tested our population of surviving scan points (see below) against the most recent HiggsTools implementation of such constraints \cite{Bahl:2022igd} and noticed very minimal differences, altogether not affecting the parametric scenarios we tested.}. 
 \item[{(ii)}] Oblique parameters\\
 The Higgs bosons in a 2HDM give contributions to the self-energies of the $W^\pm$ and $Z$ bosons.  The oblique parameters $S$, $T$ and $U$ \cite{Peskin:1990zt}, part of the aforementioned EWPOs, describe the
 deviation from the SM prediction of $S=T=U=0$. The current best-fit values (not including the recent CDF measurement of $m_W$ \cite{CDF:2022hxs}) are \cite{ParticleDataGroup:2022pth}:
 \begin{equation}
 S=-0.01\pm 0.10,\;\; T=0.03\pm 0.12,\;\; U=0.02\pm 0.11 \,.
 \end{equation}
 
If $U=0$ is taken (which is approximately true in any 2HDM Type), then the experimentally allowed ranges for $S$ and $T$ are narrowed to \cite{ParticleDataGroup:2022pth}:
 \begin{equation}
 S=0.00\pm 0.07,\;\; T=0.05\pm 0.06\,.
 \label{ST}
 \end{equation}
In our numerical results, the theoretical constraints in 1(i)--(iii) and the experimental constraints 2(ii) 
(using the ranges for $S$ and $T$ in eq.~(\ref{ST})) are respected by using 2HDMC \cite{Eriksson:2009ws}. 
If the recent measurement of $m_W$ by the CDF collaboration \cite{CDF:2022hxs} is included in the world average for $m_W$, 
 then the central values of the $S$ and $T$ parameters in eq.~(\ref{ST}) change significantly and can be accommodated in a 2HDM
 by having sizeable mass splittings among the Higgs bosons. Recent such studies have been carried out in 
\cite{Abouabid:2022lpg,Lee:2022gyf} in both NH and IH.

\item[{(iii)}] Flavour constraints\\
The parameter space of a 2HDM is also constrained by flavour observables, especially the decays of $b$-quarks (inside $B$-mesons).
The main origin of such constraints is the fact that the charged Higgs boson $H^\pm$ contributes to processes that are mediated by a $W^\pm$, leading to constraints on the
parameters $m_{H^\pm}$ and $\tan\beta$. The flavour observable that is most constraining is the rare decay $b\to s\gamma$, although $H^\pm$ contributes to numerous other 
processes (e.g., $B\overline B$ mixing). There have been many studies of flavour constraints on the parameter space of the 2HDM, see,  e.g., 
\cite{Arbey:2017gmh,Atkinson:2022pcn,Cheung:2022ndq}.
In our numerical analysis, we respect such flavour constraints by using  the publicly available code SuperIso \cite{Mahmoudi:2008tp}.
In the 2HDM Type-I, in which the couplings of $H^\pm$ to  fermions are proportional to $\cot\beta$, the constraint on $m_{H^\pm}$ is weaker with increasing $\tan\beta$.
The lowest value of  $\tan\beta$ we consider is $\tan\beta=2.5$, for which $m_{H^\pm}=140$ GeV is allowed (as can be seen in \cite{Arbey:2017gmh}).

 \end{enumerate}
  \end{enumerate}

\begin{figure}
\centering
%\begin{subfigure}
  \includegraphics[width=0.75\textwidth]{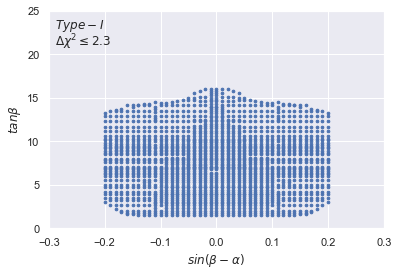}
%\end{subfigure}
\hfill
%\begin{subfigure}
    \includegraphics[width=0.75\textwidth]{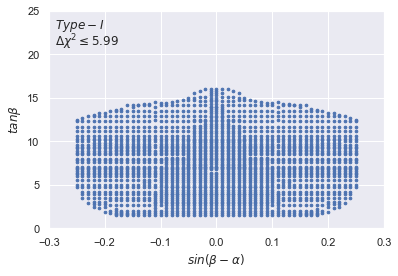}
%\end{subfigure}        
\caption{The scatter plots of the parameter space points 
surviving all theoretical constraints as well as all
experimental ones, the latter within $1\sigma$ (top) and $2\sigma$ (bottom), mapped onto the  ($\sin(\beta -\alpha ),\tan\beta$) plane for the
2HDM Type-I in IH configuration.}
\label{fig:2HDM-I}
\end{figure}

We end this section by presenting Fig.~\ref{fig:2HDM-I}, describing the parameter space of the 2HDM Type-I in IH configuration after all aforementioned constraints have been applied, as an update to Fig. 1 of Ref.~\cite{Wang:2014lta}.

\section{Results}
\label{Sec:Results}
\noindent
Based on the complete gauge-invariant set of diagrams to which those in Fig.~2 belong (hereafter, when calculating $Z^*\to l^+l^-$, we will be summing over $l=e,\mu$), 
we have computed both the integrated and differential cross section for the process $g g \rightarrow h^{0} Z^*\to h^{0}l^{+}l^{-}$ at Leading Order (LO) with MadGraph5aMC@NLO \cite{Alwall2014}, with default Parton Distribution Functions (PDFs) and corresponding factorisation/renormalisation scale\footnote{Next-to-LO (NLO) corrections in QCD exist \cite{Hasselhuhn:2016rqt,Grober:2017uho,Wang:2021rxu,Chen:2022rua,Davies:2020drs,Chen:2020gae,Degrassi:2022mro}, which -- while being sizable inclusively -- do not alter significantly the $\sqrt{\hat{s}}$  distribution (which will be of concern here). No NLO EW corrections have been computed to date.}.  Then we have used MadAnalysis \cite{Conte2013} to  analyse Monte Carlo (MC) partonic events, specifically, in looking at the  invariant mass of the system $h^{0}l^{+}l^{-}$ for a small sample of pseudo-randomly generated parameter space points, over the interval 140 GeV $<m_{A^{0}}<$ 170 GeV, {{with the mass of the charged Higgs boson 
set equal to 
the mass of the CP-odd Higgs boson (i.e., $m_{H^\pm}=m_{A^0}$).
As for the lightest Higgs boson mass ($m_{h^{0}}$), we have fixed it to 100 GeV. This value was chosen as it represents the maximum $m_{h^{0}}$ that respects both theoretical and experimental constraints. Additionally, this choice ensures that the $Z$ boson remains off-shell across all selected points.
Regarding the parameter $|\cos(\beta-\alpha)|$, as we are working within the IH scenario, it must take a value very close to $1$. Throughout our study,  $|\cos(\beta-\alpha)|$ is varied between $0.8$ and $1$.
Furthermore, to ensure theoretical and experimental feasibility for all points, we have taken $2.5<\tan\beta<10$}}. 

\begin{figure}
\includegraphics[width=1.00\textwidth]{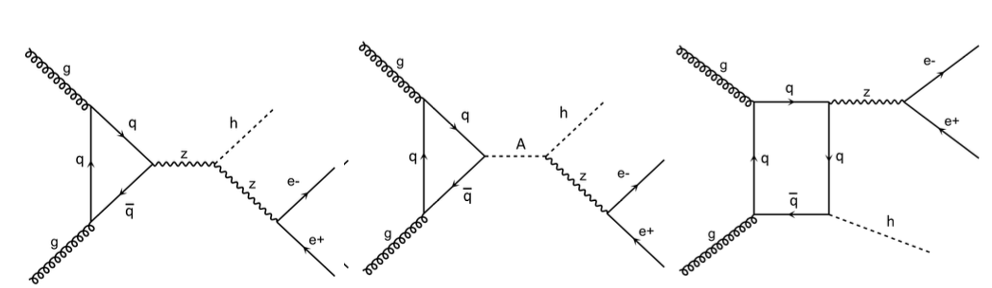}
    \label{fig:FDs}
\caption{Representative Feynman diagrams for $g g\to  h^{0} Z^*\to h^{0}e^{+}e^{-}$. (Note that the $b\bar b$ induced diagrams are negligible in the 2HDM Type-I in IH, so we ignore them throughout.) The summation is intended on all quark flavours, so that, in our BSM scenario, the first topology is dominated by $d,u,s,c$ and $b$ loops while the last two topologies are dominated by $t$ loops.}
\end{figure}

\begin{figure}
    \centering
    \includegraphics[width=12cm]{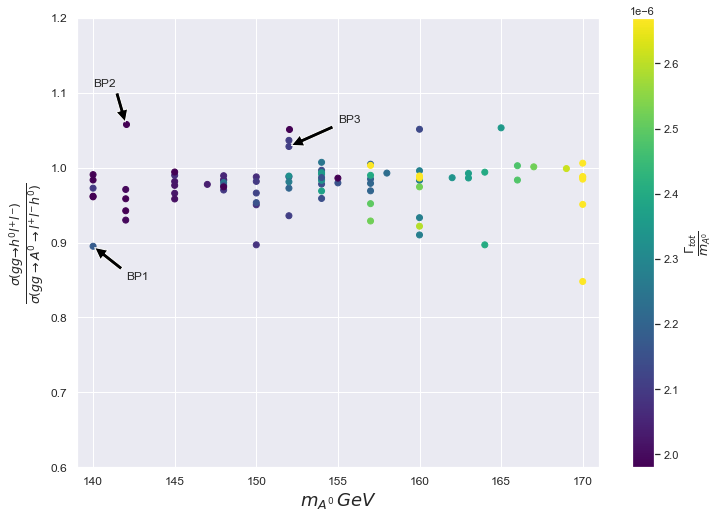}
    \caption{Values of the ratio $\frac{\sigma (gg\rightarrow h^{0}l^{+}l^{-})}{\sigma(gg\rightarrow A^{0}\rightarrow h^{0}l^{+}l^{-})}$ mapped against $m_{A^{0}}$ and colour graded against  
 $\frac{\Gamma_{A^0} }{m_{A^{0}}}$ for a selection of parameter space points surviving all constraints.
% {\textcolor{red}{The $x,y$, BP\#  and $\frac{\Gamma_{A^0}}{m_{A^{0}}}$  labels need to be much bigger}}
}
    \label{fig:integrated}
\end{figure}

In Fig.~\ref{fig:integrated} we evaluate (for some 2HDM points that respect all constraints) the ratio 
$\frac{\sigma(g g \rightarrow h^{0} Z^*\to h^{0}l^{+}l^{-})}{\sigma(g g \to A^0 \rightarrow h^{0} Z^*\to h^{0}l^{+}l^{-})}$ as a function of $m_{A^0}$,
where the denominator is the cross section corresponding to the (triangle) amplitude squared 
involving only the $A^0$ in the $s$-channel  (and in
 NWA) and the numerator is the full process, the latter including the (triangle) amplitude squared with an $s$-channel (off-shell) $Z^*$ and the amplitude squared of the box diagrams, as well as their relative interferences. For the case
 of the full process, $A^0$ is allowed to be off-shell as well. The size of the interferences depends on the value (depicted in Fig.~\ref{fig:integrated} as a colour gradient) of  the ratio $\frac{\Gamma_{A^0}}{m_{A^0}}$. One notices in Fig.~\ref{fig:integrated} that the ratio of the cross sections varies from $1.06$ to $0.85$ for the chosen points and there is only a mild correlation with $\frac{\Gamma_{A^0}}{m_{A^0}}$, which in turn highlights the counter-balancing role of the box diagrams on their own (in the positive direction) as well as that of their interference with the $A^0$ diagram (in the negative direction) on the final result
(the triangle graphs with $Z^*$ in the $s$-channel are generally negligible because of the Landau-Yan theorem \cite{Landau,Yang}). Note that this difference ($+6\% \to -15\%$) between the cross section for the full process and the factorised one in NWA  at the inclusive level in the 2HDM Type-I in IH with a light $m_{h^0}$  is somewhat higher than what was found in a previous work for the same process in the 2HDM Type-II in NH for SM-like Higgs production \cite{Accomando:2020vbo}, where  differences were confined to  the percent level (see also \cite{Kniehl:2011aa}). We further note that, although the same computation (i.e., the cross section of the full process) as the one performed here was also done in  \cite{Accomando:2020vbo}, the emerging phenomenology is rather different. This is because the size of the $A^0\to h^0 Z^*$ cross section in the 2HDM IH scenario of the present work is much larger than that in  \cite{Accomando:2020vbo} and, hence, the relative magnitude of the interferences is greater\footnote{The effects of interference between triangle and box topologies is also well known for single SM-like Higgs production in all di-boson final states \cite{Dicus:1987fk,Glover:1988fe,Glover:1988rg,vanderBij:1988fb,vanderBij:1988ac,Glover:1987nx}.}. We also emphasise that in Ref.~\cite{Accomando:2020vbo} no comparison between the factorised and full results was made.

We now select three Benchmark Points (BPs), which are labelled as BP1, BP2 and BP3 in  Fig.~\ref{fig:integrated}, for kinematical analysis. 
The ratio $\frac{\sigma (gg\rightarrow h^{0}l^{+}l^{-})}{\sigma(gg\rightarrow A^{0}\rightarrow h^{0}l^{+}l^{-})}$ is approximately $0.9$, $1.06$ and  $1.03$ for BP1, BP2 and BP3, respectively. 
 The 2HDM input parameters and cross sections for the full and factorised process for the three BPs are given in Tab.~II. 

\begin{figure}%
\vspace*{-1.0truecm}
\centering
\hspace*{-0.075truecm}\includegraphics[width=15.75cm,height=7.0cm]{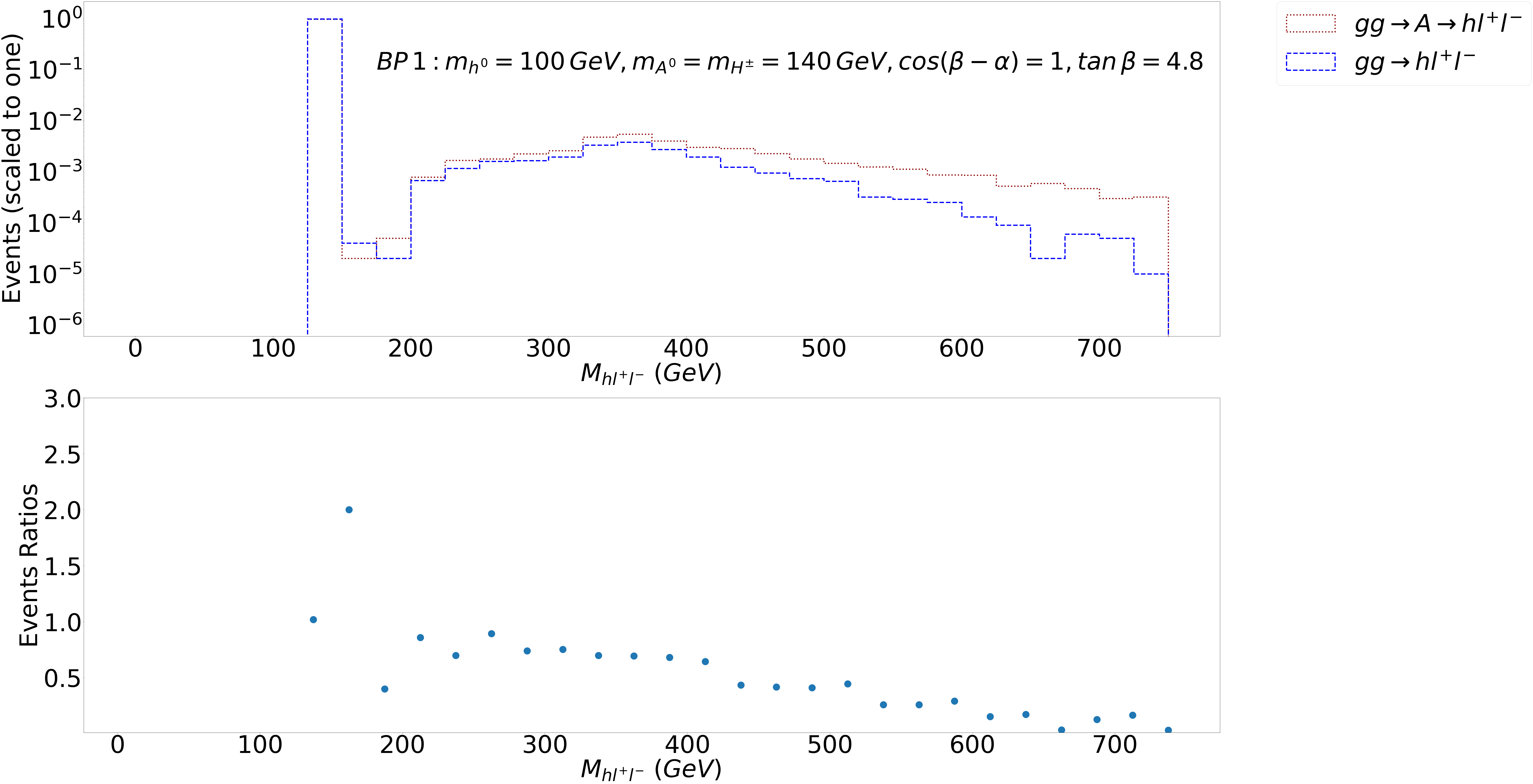}%\\
%\vspace*{0.05cm}
%\centering 
~~~~~~~~~~(a)\\
\vspace*{0.75cm}
\centering
        \includegraphics[width=15.75cm]{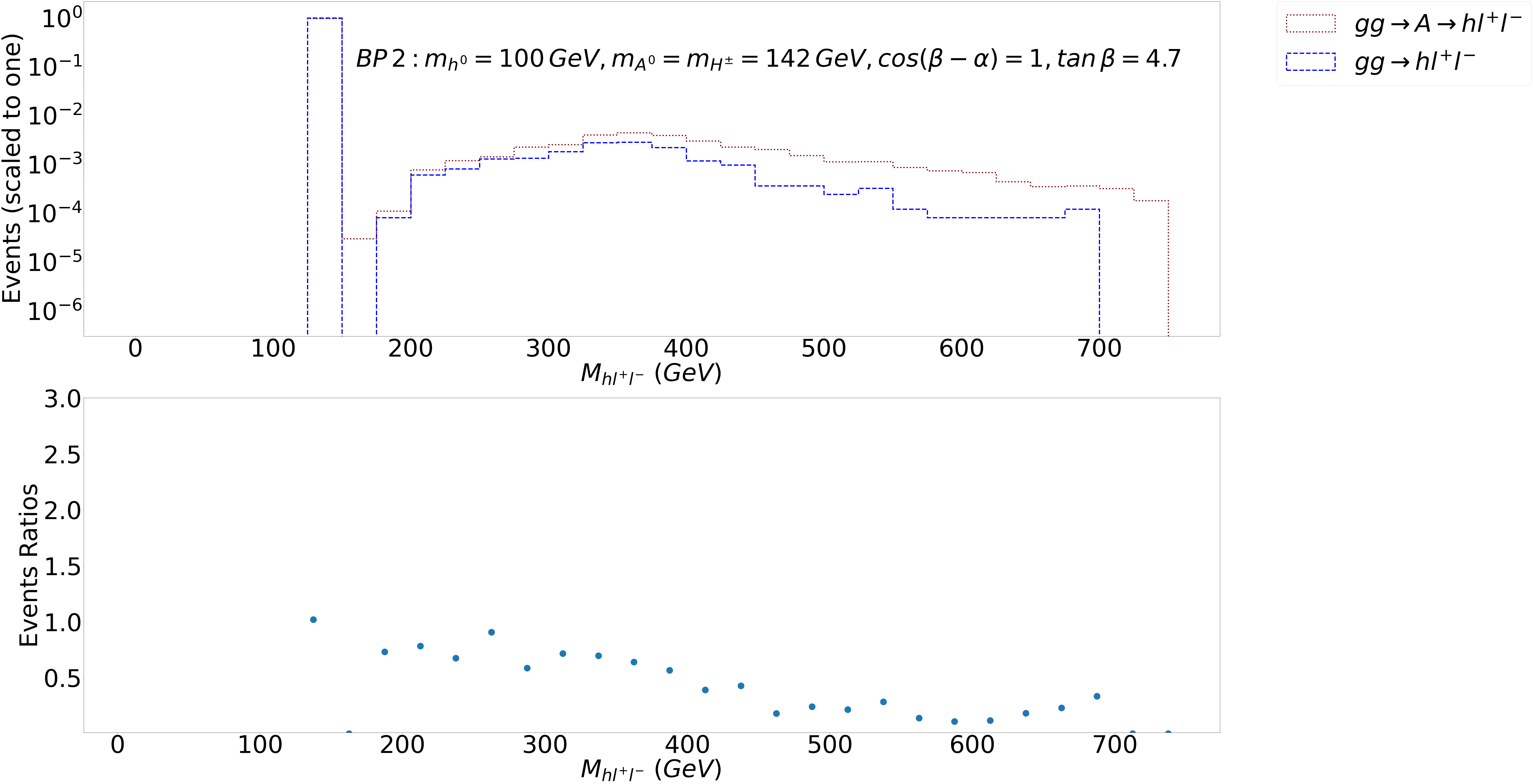}%
 \centering 
~~~~~~~(b)\\
\vspace*{0.5cm}
\centering
        \includegraphics[width=15.75cm]{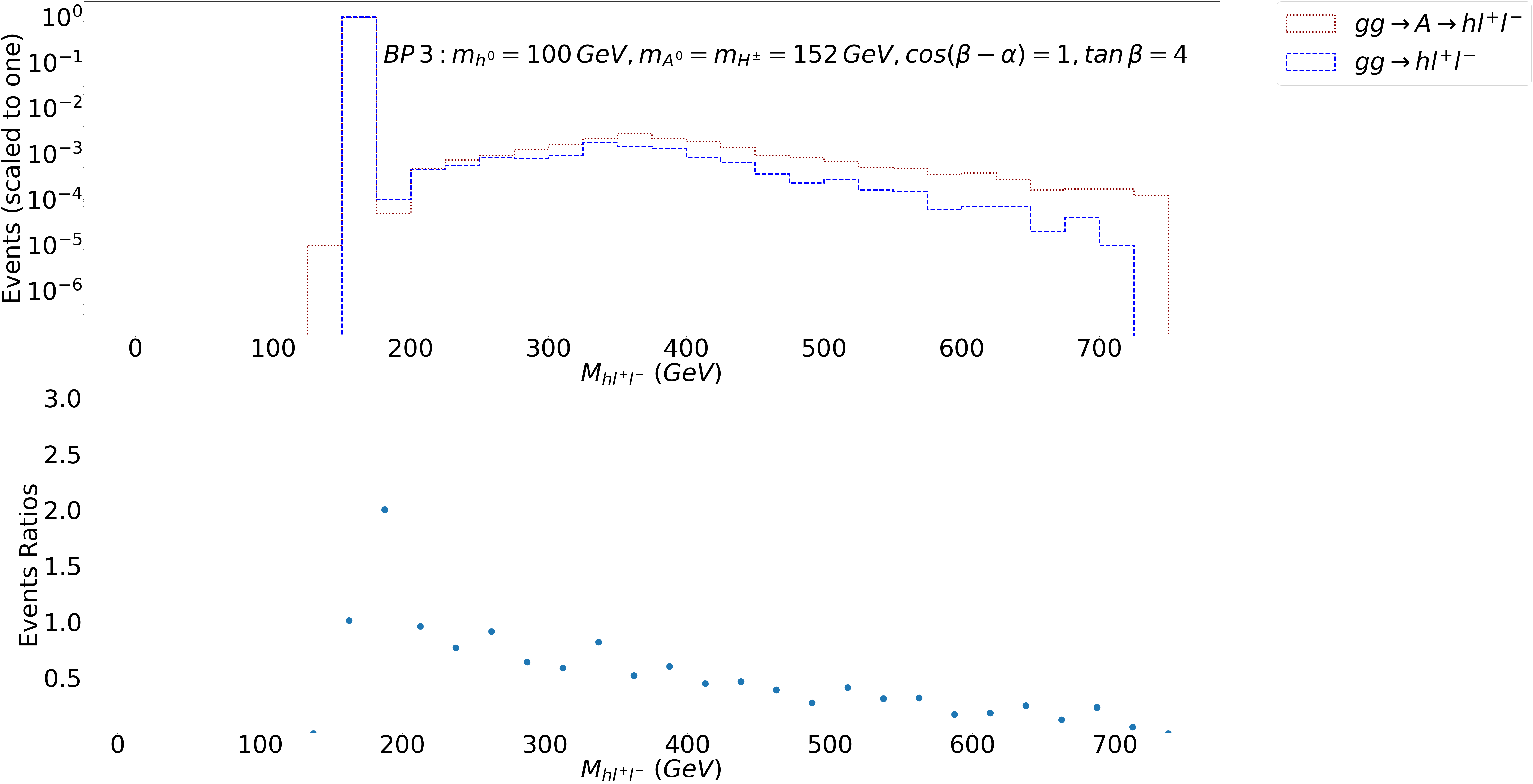}%\\
\centering 
~~~~~~~(c)
    \caption{The  $M_{h^{0} l^{+}l^{-}}$ histograms  (normalised to 1) for BP1 (a), BP2 (b) and BP3 (c) (main frames) together with their corresponding ratios (sub-frames).
% {\textcolor{red}{All axis labels and legends here should be much bigger}}
}%
    \label{fig:differential}%
\end{figure}

The invariant mass distributions for  $h^0 l^+l^-$ events of these BPs
are given in Fig.~\ref{fig:differential}. The presence of the additional diagrams in the full process 
$gg\to h^0l^+l^-$ causes a difference from the results that are obtained for the factorised process
$gg\to A^0\to h^0l^+l^-$.
In Fig.~\ref{fig:differential}, the spectra are normalised to 1 (so that we are tracking the different shapes of the two processes) and we have sampled $M_{h^0l^+l^-}$ over a bin width of 25 GeV.  In the case of BP1, the $A^0$ peak height is essentially the same for the two processes, with the factorised approximation being lower than the full result and the former taking over the latter already before the loop threshold at $\approx 2m_t$.
{{For BP2,  the $A^0$ peak height is again approximately the same for the two processes while  beyond the peak the $A^0$ process dominates (especially so starting from the aforementioned $2m_t$ value) (Notice here a strong negative interference in the full process just beyond $M_{h^0 l^+l^-}=m_{A^0}$.)}}  In BP3,  again the $A^0$ peak height is essentially the same, with the full process being larger just beyond it and subleading far beyond it, with respect to the factorised one. By looking at the ratios of the heights
of the histograms, it is clear that differences between the two approaches at the differential level can be much bigger than at the integrated level (recall Tab.~II), as the difference can be up to several tens of percent in either direction. Furthermore, such effects can have different shapes, so as to alter (unpredictably) the yield expected from a naive approach that probes the $h^0l^+l^-$ final state under the hypothesis that only the process
$gg\rightarrow A^{0}\rightarrow h^{0}l^{+}l^{-}$ contributes, as is customarily assumed in typical experimental analyses. However, as can be seen from Fig.~\ref{fig:differential}, the fraction of events that is away from the $A^0$ peak is very small
(to be quantified below) and thus we expect that such a difference in the differential distributions (in the event of a discovery of $A^0\to h^0Z^*$) would only (possibly) start to be manifest with much more integrated luminosity.
In fact, the actual size of such effects in an experimental analysis depends on the 
way the invariant mass of $b\bar bl^+l^-$   is sampled (here we assume the predominant $h^0\to b\bar b$ decay and that the value of $m_{h^0}$ can be reconstructed in actual LHC analyses.)

{%\textcolor{blue}
{In existing searches for $A^0$ narrow resonances \cite{1911.03781}, whereby a tentative hypothesis is made for the value of $m_{A^0}$, so that selection cuts can be optimised accordingly,  and the $M_{h^0l^+l^-}$  distribution is sampled in narrow bins,  e.g., of 25 GeV (a typical resolution of the $b\bar bl^+l^-$ system), 
 the effect of the above results for the three BPs would correspond to rescaling the naive expectation stemming from the  process $gg\rightarrow A^{0}\rightarrow h^{0}l^{+}l^{-}$ by the amounts $-10\%$, $+6\%$ and  $+3\%$ for BP1, BP2 and BP3, respectively (i.e., by the corrections at inclusive level). This is clear from Fig.~\ref{fig:differential}, as most of the $A^0\to h^0Z^*$ events fall in one bin that contains $M_{A^0}$. Indeed, the percentage of the differential cross section for the full process in Fig.~\ref{fig:differential} in the 25 GeV bin centred around $m_{A^0}$ is $97.6(98.2)[98.8]\%$, so that only a very small part of  it  is found in the remaining bins. 
}}

{%\textcolor{blue}
{However, if a non-resonant search is performed, thereby using for the significance calculation a much wider expanse in such an invariant mass (i.e., with no tentative assumption made on  $m_{A^0}$  or $\Gamma_{A^0}$), and  the mass intervals used for sampling did not include the $A^0$ peak (e.g., because `blind' selection criteria would eliminate it from the candidate signal sample or such a mass interval is chosen elsewhere from the peak region), overall effects could be drastically different. For example,  if one sampled the two processes 
${gg\rightarrow h^{0}l^{+}l^{-}}$ and 
${gg\rightarrow A^{0}\rightarrow h^{0}l^{+}l^{-}}$ over the mass intervals 200 GeV $< M_{h^0l^+l^-}<$ 400(600)[750] GeV,  differences between the two would be 
$-71(-60.4)[-57.7]\%$  for BP1, 
$-63(-51.4)[-49.7]\%$  for BP2 and $-64(-55.9)[-54.2]\%$  for BP3\footnote{Note that there are  negative interferences in the mass region between $m_{A^0}$ and $m_{h^0}+m_Z$ which  have deliberately not been  included in our $M_{h^0l^+l^-}$  sampling and could make such differences even more dramatic.}. Clearly, such significant effects should not be surprising: they are indeed consistent with the inclusive cross section results since these bins only contain a small fraction of the total number of $h^0 l^+l^-$ events. However, we suggest that they could start to be evident at high integrated luminosity. This may happen under two circumstances. On the hand, they could emerge following a presumed discovery of the resonant $A^0$ state (at a lower integrated luminosity), whereby one could use the high mass regions of the $M_{h^0l^+l^-}$ distribution to extract its properties, notably, around $2m_t$. On the other hand, in the absence of such a discovery, they could simply manifest themselves as a broad excess above the SM preditions, which may be retroactively be used to search for the $A^0$ state, produced resonantly outside of the sampled  $M_{h^0l^+l^-}$ interval.
}}

\begin{table}[!t]
    \centering
    \begin{tabular}{ |p{1.1cm}|p{1.2cm}|p{1.5cm}|p{4.5cm}|p{3.2cm}|}
 \hline
  BP&$m_{A^{0}}$ & $\tan\beta$&  $\sigma (gg\to A^{0}\to h^{0}l^{+}l^{-})$&$\sigma (g g \rightarrow h^{0} l^{+} l^{-})$ \\ 
  \hline
 1&140 &  4.8& 0.01296&0.01168\\\hline
  2&142 & 4.7&  0.01458&0.01543\\
  \hline
  3&152 & 4&0.03859&0.03964 \\
  
  \hline
\end{tabular}
    \label{tab:BPs}
\vspace*{0.5cm}
    \caption{The input values for the parameters defining our three BPs and corresponding  cross sections in pb, for all of these we have $\cos(\beta -\alpha) =1$, $m_{h^0}=100$ GeV and $m_{H^{\pm }}=m_{A^0}$. (Also recall that $m_{H^0}=$ 125 GeV.)}
\end{table}

\section{Conclusions}  
\label{Sec:Summary}
\noindent
In summary, we have described the phenomenology of the $gg\to h^0Z^*(\to l^+l^-)$  process at the LHC, in the context of the 2HDM Type-I (within NFC)  in the IH scenario, i.e., with  $m_{h^0}< m_{H^0}=125$ GeV. We have shown that sizable differences can exist between the naive approach wherein the above process is factorised with the $A^0$ state in NWA 
(i.e., $gg\to A^0\to  h^0Z^*$) 
and the full process in which such a Higgs state can be off-shell and also accounting for all other Feynman diagram structures entering it (and corresponding interferences):  i.e.,  $gg\to Z^*\to h^0Z^*$ (through one-loop triangle diagrams) and $gg\to h^0Z^*$ (through one-loop box diagrams). Such  differences vary significantly depending on the kind of analysis which is deployed to search for the CP-odd Higgs state, whether resonant or non-resonant, but in both cases amounting to up to several tens of percent  at the differential level while at the integrated level rates are never beyond 10\% or so. Hence, in the circumstances, a redefinition of the signal is necessary, so as to also capture all such effects entering the cross section involving $A^0$ production and decay, 
alongside the use of  a suitable computational framework.  

Indeed, in order to aid the experimental pursuit of the effects described here, we have  released in this paper several BPs representative of parameter space configurations enabling sensitivity to what are the most effective (i.e., resonant, where the majority of the signal would manifest itself) searches for  $h^0Z^*(\to l^+l^-)$ production and decay as well as  potentially visible tail effects (where only a minority of the signal would be present) from such a process appearing in non-resonant searches (i.e., those not specifically optimised to the model and/or process considered here). Finally, we will eventually make the computing tools developed here also available in the public domain.

\noindent
\section*{Acknowledgements}
\noindent
SA acknowledges the use of the IRIDIS High Performance Computing Facility, and associated support services at the University of Southampton. SA acknowledges
support from a scholarship of the Imam Mohammad Ibn Saud Islamic University.
AA and SM are funded in part through the STFC CG ST/L000296/1.
SM is funded in part through the NExT Institute. We thank Souad Semlali for useful discussions.

\newpage
\pagenumbering{gobble}

\begin{thebibliography}{99}

%\cite{Aad:2012tfa}
\bibitem{Aad:2012tfa} 
  G.~Aad {\it et al.} [ATLAS Collaboration],
  %``Observation of a new particle in the search for the Standard Model Higgs boson with the ATLAS detector at the LHC,''
  Phys.\ Lett.\ B {\bf 716}, 1 (2012)
%  %doi:10.1016/j.physletb.2012.08.020
  [arXiv:1207.7214 [hep-ex]].
  %%CITATION = %doi:10.1016/j.physletb.2012.08.020;%%
  %6047 citations counted in INSPIRE as of 17 May 2016

%\cite{Chatrchyan:2012xdj}
\bibitem{Chatrchyan:2012xdj} 
  S.~Chatrchyan {\it et al.} [CMS Collaboration],
  %``Observation of a new boson at a mass of 125 GeV with the CMS experiment at the LHC,''
  Phys.\ Lett.\ B {\bf 716}, 30 (2012)
%  %doi:10.1016/j.physletb.2012.08.021
  [arXiv:1207.7235 [hep-ex]].
  %%CITATION = %doi:10.1016/j.physletb.2012.08.021;%%
  %5904 citations counted in INSPIRE as of 17 May 2016

%%%%%%%%%%%%%%%%%%%%%%%%%%%%%%%%%%
%%\cite{ATLAS:2018kot}
%\bibitem{ATLAS:2018kot}
%M.~Aaboud \textit{et al.} [ATLAS],
%%``Observation of $H \rightarrow b\bar{b}$ decays and $VH$ production with the ATLAS detector,''
%Phys. Lett. B \textbf{786}, 59-86 (2018)
%%%doi:10.1016/j.physletb.2018.09.013
%[arXiv:1808.08238 [hep-ex]].
%%371 citations counted in INSPIRE as of 06 Jan 2022
%
%%\cite{CMS:RunII}
%\bibitem{CMS:RunII}
%[CERN CMS Collaboration], CMS PAS HIG-19-005.
%
%
%%\cite{ATLAS:RunII}
%\bibitem{ATLAS:RunII}
%[CERN ATLAS Collaboration], ATLAS-CONF-2021-053.
%%%%%%%%%%%%%%%%%%%%%%%%%%%%%%%%%%

%\cite{Lee:1973iz}
\bibitem{Lee:1973iz}
T.~D.~Lee,
%``A Theory of Spontaneous T Violation,''
Phys. Rev. D \textbf{8}, 1226-1239 (1973).
%%doi:10.1103/PhysRevD.8.1226
%1482 citations counted in INSPIRE as of 14 Aug 2022

%\cite{Gunion:1989we}
\bibitem{Gunion:1989we}
J.~F.~Gunion, H.~E.~Haber, G.~L.~Kane and S.~Dawson,
%``The Higgs Hunter's Guide,''
Front. Phys. \textbf{80}, 1-404 (2000).
%1219 citations counted in INSPIRE as of 14 Aug 2022


%\cite{Branco:2011iw}
\bibitem{Branco:2011iw} 
  G.~C.~Branco, P.~M.~Ferreira, L.~Lavoura, M.~N.~Rebelo, M.~Sher and J.~P.~Silva,
  %``Theory and phenomenology of two-Higgs-doublet models,''
  Phys.\ Rept.\  {\bf 516}, 1 (2012)
%  %doi:10.1016/j.physrep.2012.02.002
  [arXiv:1106.0034 [hep-ph]].
  %%CITATION = %doi:10.1016/j.physrep.2012.02.002;%%
  %1292 citations counted in INSPIRE as of 29 Aug 2018

%\cite{Wang:2022yhm}
\bibitem{Wang:2022yhm}
L.~Wang, J.~M.~Yang and Y.~Zhang,
%``Two-Higgs-doublet models in light of current experiments: a brief review,''
[arXiv:2203.07244 [hep-ph]].
%5 citations counted in INSPIRE as of 10 Aug 2022

%\cite{Akeroyd:2023kek}
\bibitem{Akeroyd:2023kek}
A.~G.~Akeroyd, S.~Alanazi and S.~Moretti,
%``The decay A $^{0}$ \textrightarrow{} h $^{0}$ Z(\ensuremath{*}) in the inverted hierarchy scenario and its detection prospects at the large hadron collider,''
J. Phys. G \textbf{50} (2023) no.9, 095001
% %doi:10.1088/1361-6471/ace3e1
[arXiv:2301.00728 [hep-ph]].
%1 citations counted in INSPIRE as of 12 Jan 2024

%\cite{Accomando:2020vbo}
\bibitem{Accomando:2020vbo}
E.~Accomando, M.~Chapman, A.~Maury and S.~Moretti,
%``Below-threshold CP-odd Higgs boson search via $A\rightarrow Z^*h$ at the LHC,''
Phys. Lett. B \textbf{818}, 136342 (2021)
%%doi:10.1016/j.physletb.2021.136342
[arXiv:2002.07038 [hep-ph]].
%1 citations counted in INSPIRE as of 10 Aug 2022

%%%%%%%%%%%%%%%%%%%%%%%%%%%%%%%%%%%%
%%\cite{Djouadi:2005gj}
%\bibitem{Djouadi:2005gj}
%A.~Djouadi,
%%``The Anatomy of electro-weak symmetry breaking. II. The Higgs bosons in the minimal supersymmetric model,''
%Phys. Rept. \textbf{459}, 1-241 (2008)
%%%doi:10.1016/j.physrep.2007.10.005
%[arXiv:hep-ph/0503173 [hep-ph]].
%%1507 citations counted in INSPIRE as of 10 Aug 2022
%%%%%%%%%%%%%%%%%%%%%%%%%%%%%%%%%%%%

%\cite{Glashow:1976nt}
\bibitem{Glashow:1976nt} 
  S.~L.~Glashow and S.~Weinberg,
  %``Natural Conservation Laws for Neutral Currents,''
  Phys.\ Rev.\ D {\bf 15}, 1958 (1977);
%  %doi:10.1103/PhysRevD.15.1958
  %%CITATION = %doi:10.1103/PhysRevD.15.1958;%%
  %1650 citations counted in INSPIRE as of 06 Sep 2018
%\cite{Paschos:1976ay}
%\bibitem{Paschos:1976ay} 
  E.~A.~Paschos,
  %``Diagonal Neutral Currents,''
  Phys.\ Rev.\ D {\bf 15}, 1966 (1977).
%  %doi:10.1103/PhysRevD.15.1966
  %%CITATION = %doi:10.1103/PhysRevD.15.1966;%%
  %481 citations counted in INSPIRE as of 06 Sep 2018

%\cite{Barger} 
\bibitem{Barger} 
  V.~D.~Barger, J.~L.~Hewett and R.~J.~N.~Phillips,
  %``New Constraints on the Charged Higgs Sector in Two Higgs Doublet Models,''
  Phys.\ Rev.\ D {\bf 41}, 3421 (1990).
  %%CITATION = %doi:10.1103/PhysRevD.41.3421;%%
  %445 citations counted in INSPIRE as of 28 Jan 2016

%\cite{Bechtle:2020pkv}
\bibitem{Bechtle:2020pkv}
P.~Bechtle, D.~Dercks, S.~Heinemeyer, T.~Klingl, T.~Stefaniak, G.~Weiglein and J.~Wittbrodt,
%``HiggsBounds-5: Testing Higgs Sectors in the LHC 13 TeV Era,''
Eur. Phys. J. C \textbf{80}, no.12, 1211 (2020)
%%doi:10.1140/epjc/s10052-020-08557-9
[arXiv:2006.06007 [hep-ph]].
%118 citations counted in INSPIRE as of 14 Aug 2022

%\cite{Bechtle:2020uwn}
\bibitem{Bechtle:2020uwn}
P.~Bechtle, S.~Heinemeyer, T.~Klingl, T.~Stefaniak, G.~Weiglein and J.~Wittbrodt,
%``HiggsSignals-2: Probing new physics with precision Higgs measurements in the LHC 13 TeV era,''
Eur. Phys. J. C \textbf{81}, no.2, 145 (2021)
%%doi:10.1140/epjc/s10052-021-08942-y
[arXiv:2012.09197 [hep-ph]].
%92 citations counted in INSPIRE as of 14 Aug 2022

%\cite{Bahl:2022igd}
\bibitem{Bahl:2022igd}
H.~Bahl, T.~Biek\"otter, S.~Heinemeyer, C.~Li, S.~Paasch, G.~Weiglein and J.~Wittbrodt,
%``HiggsTools: BSM scalar phenomenology with new versions of HiggsBounds and HiggsSignals,''
Comput. Phys. Commun. \textbf{291}, 108803 (2023)
%doi:10.1016/j.cpc.2023.108803
[arXiv:2210.09332 [hep-ph]].
%42 citations counted in INSPIRE as of 26 Jan 2024

%\cite{Peskin:1990zt}
\bibitem{Peskin:1990zt}
M.~E.~Peskin and T.~Takeuchi,
%``A New constraint on a strongly interacting Higgs sector,''
Phys. Rev. Lett. \textbf{65}, 964-967 (1990).
%%doi:10.1103/PhysRevLett.65.964
%2009 citations counted in INSPIRE as of 23 Jan 2022

%\cite{CDF:2022hxs}
\bibitem{CDF:2022hxs}
T.~Aaltonen \textit{et al.} [CDF Collaboration],
%``High-precision measurement of the $W$          boson mass with the CDF II detector,''
Science \textbf{376}, no.6589, 170-176 (2022).
%%doi:10.1126/science.abk1781
%285 citations counted in INSPIRE as of 31 Dec 2022

%\cite{ParticleDataGroup:2022pth}
\bibitem{ParticleDataGroup:2022pth}
R.~L.~Workman \textit{et al.} [Particle Data Group],
%``Review of Particle Physics,''
PTEP \textbf{2022}, 083C01 (2022).
%%doi:10.1093/ptep/ptac097
%424 citations counted in INSPIRE as of 31 Dec 2022


%\cite{Eriksson:2009ws}
\bibitem{Eriksson:2009ws}
D.~Eriksson, J.~Rathsman and O.~St{\aa}l,
%``2HDMC: Two-Higgs-Doublet Model Calculator Physics and Manual,''
Comput. Phys. Commun. \textbf{181}, 189-205 (2010)
%%doi:10.1016/j.cpc.2009.09.011
[arXiv:0902.0851 [hep-ph]].
%426 citations counted in INSPIRE as of 14 Aug 2022

%\cite{Abouabid:2022lpg}
\bibitem{Abouabid:2022lpg}
H.~Abouabid, A.~Arhrib, R.~Benbrik, M.~Krab and M.~Ouchemhou,
%``Is the new CDF $M_W$ measurement consistent with the two higgs doublet model?,''
[arXiv:2204.12018 [hep-ph]].
%23 citations counted in INSPIRE as of 27 Aug 2022

%\cite{Lee:2022gyf}
\bibitem{Lee:2022gyf}
S.~Lee, K.~Cheung, J.~Kim, C.~T.~Lu and J.~Song,
%``Status of the two-Higgs-doublet model in light of the CDF $m_W$ measurement,''
[arXiv:2204.10338 [hep-ph]].
%34 citations counted in INSPIRE as of 27 Aug 2022


%\cite{Arbey:2017gmh}
\bibitem{Arbey:2017gmh} 
  A.~Arbey, F.~Mahmoudi, O.~Stal and T.~Stefaniak,
  %``Status of the Charged Higgs Boson in Two Higgs Doublet Models,''
  Eur.\ Phys.\ J.\ C {\bf 78}, no. 3, 182 (2018)
%  %doi:10.1140/epjc/s10052-018-5651-1
  [arXiv:1706.07414 [hep-ph]].
  %%CITATION = %doi:10.1140/epjc/s10052-018-5651-1;%%
  %28 citations counted in INSPIRE as of 30 Aug 2018


%\cite{Cheung:2022ndq}
\bibitem{Cheung:2022ndq}
K.~Cheung, A.~Jueid, J.~Kim, S.~Lee, C.~T.~Lu and J.~Song,
%``Comprehensive study of the light charged Higgs boson in the type-I two-Higgs-doublet model,''
Phys. Rev. D \textbf{105}, no.9, 095044 (2022)
%%doi:10.1103/PhysRevD.105.095044
[arXiv:2201.06890 [hep-ph]].
%12 citations counted in INSPIRE as of 27 Aug 2022


%\cite{Atkinson:2022pcn}
\bibitem{Atkinson:2022pcn}
O.~Atkinson, M.~Black, C.~Englert, A.~Lenz, A.~Rusov and J.~Wynne,
%``The Flavourful Present and Future of 2HDMs at the Collider Energy Frontier,''
[arXiv:2202.08807 [hep-ph]].
%10 citations counted in INSPIRE as of 27 Aug 2022

%\cite{Mahmoudi:2008tp}
\bibitem{Mahmoudi:2008tp}
F.~Mahmoudi,
%``SuperIso v2.3: A Program for calculating flavor physics observables in Supersymmetry,''
Comput. Phys. Commun. \textbf{180}, 1579-1613 (2009)
%%doi:10.1016/j.cpc.2009.02.017
[arXiv:0808.3144 [hep-ph]].
%392 citations counted in INSPIRE as of 27 Aug 2022

%\cite{Wang:2014lta}
\bibitem{Wang:2014lta}
L.~Wang and X.~F.~Han,
%``Study of the heavy CP-even Higgs with mass 125 GeV in two-Higgs-doublet models at the LHC and ILC,''
JHEP \textbf{11}, 085 (2014)
%doi:10.1007/JHEP11(2014)085
[arXiv:1404.7437 [hep-ph]].

%\cite{Alwall2014}
\bibitem{Alwall2014}
J.~Alwall, R.~Frederix, S.~Frixione, V.~Hirschi, F.~Maltoni, O.~Mattelaer, H.~S.~Shao, T.~Stelzer, P.~Torrielli and M.~Zaro,
%``The automated computation of tree-level and next-to-leading order differential cross sections, and their matching to parton shower simulations,''
JHEP \textbf{07}, 079 (2014)
%doi:10.1007/JHEP07(2014)079
[arXiv:1405.0301 [hep-ph]].
%8339 citations counted in INSPIRE as of 02 Apr 2024

%\cite{Hasselhuhn:2016rqt}
\bibitem{Hasselhuhn:2016rqt}
A.~Hasselhuhn, T.~Luthe and M.~Steinhauser,
%``On top quark mass effects to $gg\to ZH$ at NLO,''
JHEP \textbf{01}, 073 (2017)
%doi:10.1007/JHEP01(2017)073
[arXiv:1611.05881 [hep-ph]].
%31 citations counted in INSPIRE as of 02 Apr 2024

%\cite{Grober:2017uho}
\bibitem{Grober:2017uho}
R.~Gr\"ober, A.~Maier and T.~Rauh,
%``Reconstruction of top-quark mass effects in Higgs pair production and other gluon-fusion processes,''
JHEP \textbf{03}, 020 (2018)
%doi:10.1007/JHEP03(2018)020
[arXiv:1709.07799 [hep-ph]].
%61 citations counted in INSPIRE as of 02 Apr 2024

%\cite{Davies:2020drs}
\bibitem{Davies:2020drs}
J.~Davies, G.~Mishima and M.~Steinhauser,
%``Virtual corrections to $gg\to ZH$ in the high-energy and large-$m_t$ limits,''
JHEP \textbf{03}, 034 (2021)
%doi:10.1007/JHEP03(2021)034
[arXiv:2011.12314 [hep-ph]].
%26 citations counted in INSPIRE as of 02 Apr 2024

%\cite{Chen:2020gae}
\bibitem{Chen:2020gae}
L.~Chen, G.~Heinrich, S.~P.~Jones, M.~Kerner, J.~Klappert and J.~Schlenk,
%``$ZH$ production in gluon fusion: two-loop amplitudes with full top quark mass dependence,''
JHEP \textbf{03}, 125 (2021)
%doi:10.1007/JHEP03(2021)125
[arXiv:2011.12325 [hep-ph]].
%31 citations counted in INSPIRE as of 02 Apr 2024

%\cite{Wang:2021rxu}
\bibitem{Wang:2021rxu}
G.~Wang, X.~Xu, Y.~Xu and L.~L.~Yang,
%``Next-to-leading order corrections for gg \textrightarrow{} ZH with top quark mass dependence,''
Phys. Lett. B \textbf{829}, 137087 (2022)
%doi:10.1016/j.physletb.2022.137087
[arXiv:2107.08206 [hep-ph]].
%19 citations counted in INSPIRE as of 02 Apr 2024

%\cite{Chen:2022rua}
\bibitem{Chen:2022rua}
L.~Chen, J.~Davies, G.~Heinrich, S.~P.~Jones, M.~Kerner, G.~Mishima, J.~Schlenk and M.~Steinhauser,
%``ZH production in gluon fusion at NLO in QCD,''
JHEP \textbf{08}, 056 (2022)
%doi:10.1007/JHEP08(2022)056
[arXiv:2204.05225 [hep-ph]].
%17 citations counted in INSPIRE as of 02 Apr 2024

%\cite{Degrassi:2022mro}
\bibitem{Degrassi:2022mro}
G.~Degrassi, R.~Gr\"ober, M.~Vitti and X.~Zhao,
%``On the NLO QCD corrections to gluon-initiated ZH production,''
JHEP \textbf{08}, 009 (2022)
%doi:10.1007/JHEP08(2022)009
[arXiv:2205.02769 [hep-ph]].
%14 citations counted in INSPIRE as of 02 Apr 2024

%\cite{Conte2013}
\bibitem{Conte2013}
E.~Conte, B.~Fuks and G.~Serret,
%``MadAnalysis 5, A User-Friendly Framework for Collider Phenomenology,''
Comput. Phys. Commun. \textbf{184}, 222-256 (2013)
%doi:10.1016/j.cpc.2012.09.009
[arXiv:1206.1599 [hep-ph]].
%597 citations counted in INSPIRE as of 02 Apr 2024

\bibitem{Landau} L.D. Landau, Dokl. Akad. Nauk SSSR 60, 207 (1948) 207.

\bibitem{Yang}  C.~N. Yang, Phys. Rev. 77, 242 (1950).

%\cite{Kniehl:2011aa}
\bibitem{Kniehl:2011aa}
B.~A.~Kniehl and C.~P.~Palisoc,
%``Associated production of Z and neutral Higgs bosons at the CERN Large Hadron Collider,''
Phys. Rev. D \textbf{85}, 075027 (2012) 
%doi:10.1103/PhysRevD.85.075027
[arXiv:1112.1575 [hep-ph]].
%15 citations counted in INSPIRE as of 02 Apr 2024

%\cite{Dicus:1987fk}
\bibitem{Dicus:1987fk}
D.~A.~Dicus and S.~S.~D.~Willenbrock,
%``Photon Pair Production and the Intermediate Mass Higgs Boson,''
Phys. Rev. D \textbf{37}, 1801 (1988).
%doi:10.1103/PhysRevD.37.1801
%127 citations counted in INSPIRE as of 09 Apr 2024

%\cite{Glover:1988fe}
\bibitem{Glover:1988fe}
E.~W.~N.~Glover and J.~J.~van der Bij,
%``VECTOR BOSON PAIR PRODUCTION VIA GLUON FUSION,''
Phys. Lett. B \textbf{219}, 488-492 (1989).
%doi:10.1016/0370-2693(89)91099-X
%145 citations counted in INSPIRE as of 29 Feb 2024

%\cite{Glover:1988rg}
\bibitem{Glover:1988rg}
E.~W.~N.~Glover and J.~J.~van der Bij,
%``Z BOSON PAIR PRODUCTION VIA GLUON FUSION,''
Nucl. Phys. B \textbf{321}, 561-590 (1989).
%doi:10.1016/0550-3213(89)90262-9
%243 citations counted in INSPIRE as of 29 Feb 2024

%\cite{vanderBij:1988fb}
\bibitem{vanderBij:1988fb}
J.~J.~van der Bij and E.~W.~N.~Glover,
%``PHOTON Z BOSON PAIR PRODUCTION VIA GLUON FUSION,''
Phys. Lett. B \textbf{206}, 701-704 (1988).
%doi:10.1016/0370-2693(88)90722-8
%58 citations counted in INSPIRE as of 29 Feb 2024

%\cite{vanderBij:1988ac}
\bibitem{vanderBij:1988ac}
J.~J.~van der Bij and E.~W.~N.~Glover,
%``$Z$ Boson Production and Decay via Gluons,''
Nucl. Phys. B \textbf{313}, 237-257 (1989).
%doi:10.1016/0550-3213(89)90317-9
%82 citations counted in INSPIRE as of 09 Apr 2024

%\cite{Glover:1987nx}
\bibitem{Glover:1987nx}
E.~W.~N.~Glover and J.~J.~van der Bij,
%``HIGGS BOSON PAIR PRODUCTION VIA GLUON FUSION,''
Nucl. Phys. B \textbf{309}, 282-294 (1988).
%doi:10.1016/0550-3213(88)90083-1
%367 citations counted in INSPIRE as of 09 Apr 2024

\bibitem{1911.03781}
S.~Chatrchyan {\it et al.} [CMS Collaboration], axXiv:1911.03781 [hep-ex].




\end{thebibliography}
\end{document}